\documentclass[jkps,preprint,fleqn,showpacs,showkeys]{revtex4}
\usepackage{graphicx}
\usepackage{epstopdf}
\usepackage{dcolumn}
\usepackage{bm}
\usepackage{amssymb}
\usepackage{amsmath}
\begin{document}
\setcounter{page}{0}
\title{ Anisotropic two-dimensional RF-dressed potentials for ultracold atoms}
\author{Arijit \surname{Chakraborty}}
\email[E-mail me at: ]{carijit@rrcat.gov.in}
\author{Satya Ram \surname{Mishra}}
\email[E-mail me at: ]{srm@rrcat.gov.in}
\affiliation{Laser Physics Applications Section,\\ Raja Ramanna Center for Advanced Technology,\\Indore-452013, India.}
\begin{abstract}
In this work, the RF-dressed potentials generated using a static magnetic field of a quadrupole trap and various radio frequency (RF) fields, have been theoretically investigated for trapping and manipulations of cold atoms in a two-dimensional (2D) geometry. It is shown that, in this scheme, the RF fields of some particular polarizations and phases can give rise to some novel static and time-dependent anisotropic two-dimensional potentials which are useful to trap and manipulate the cold atoms in the 2D geometry. The generated time-dependent 2D potentials, rotating on the circular ring, can be used for the controlled rotation and oscillation of the cold atom cloud on the circular ring path. The ultracold atoms trapped in these potentials may be used to investigate the interesting physics phenomena such as tunnelling and super-fluidity.
\end{abstract}
\maketitle
\section{Introduction}
\hspace*{5pt}
Trapping and manipulation of ultracold atoms is of significant importance due to applications of cold atoms in several areas such as study of quantum degenerate gases \cite{EnricoFermiCourse:1999}, quantum information processing \cite{Cirac:1995,Garcia-Ripoll:2003}, atom interferometry \cite{Berman:book,Andrews:1997}, precession measurement \cite{Segal:thesis,Canuel:2006} etc. An atom with finite value of its magnetic dipole moment can be trapped in an inhomogeneous static magnetic field at the position of the minimum potential energy \cite{EnricoFermiCourse:1999}. The potential energy of an atom in a magnetic field B is given as $V=m_Fg_F\mu_BB$, where $m_F$ is the Zeeman hyperfine quantum number for the state with hyperfine number $F$, $g_F$ is the Lande-g factor for the hyperfine state, and $\mu_B$ is the Bohr magneton \cite{CJFoot:book}. Depending on the sign of the product $m_Fg_F$, the atom can seek either the minimum of the field or the maximum of the field to achieve the minimum potential energy. The states with  $m_Fg_F>0$ are called low field seeking states, whereas the states with $m_Fg_F<0$ are called high field seeking states. Since Maxwell's equation $(\nabla.B=0)$ together with Earnshaw's theorem denies the existence of the maximum of the magnetic field in free space, the trapping with static magnetic field is possible only for atoms in low field seeking states at the minimum of the magnetic field. Therefore, the trapping geometries achievable using the static magnetic field alone remain limited. To overcome this limitation, Zobay and Garraway \cite{Zobay:2001,Zobay:2004} proposed the use of the radio frequency (RF) field in the presence of the static magnetic field to generate the adiabatic dressed state potentials for an atom. These potentials, referred as `RF-dressed potentials' in short, offer extensive control over potential landscapes and trapping geometries through RF field parameters such as amplitude, polarization, frequency and phase. This can meet the enormous demand of versatility required for trapping and manipulation of the cold atoms for various purposes ranging from the study of fundamental physics to matter wave interferometry based precision measurements \cite{Hofferberth:2007,Courteille:2006,Berman:book}. These potentials show good flexibility and versatility for trapping and manipulation of cold atoms on miniaturized scale as well \cite{Lesanovsky:2006:73,Lesanovsky:2006:74,Folman:2002}. It turns out that these dressed state potentials are inherently inhomogeneous due to position dependent direction and magnitude of the static magnetic field which controls the Rabi frequency. Since the first proposal by Zobay and Garraway \cite{Zobay:2001}, the atom trapping in these RF-dressed potentials has been studied theoretically as well as experimentally by several groups world wide \cite{Colombe:2004,Schumm:2005,Morizot:2006,White:2006,Lesanovsky:2006:74,Courteille:2006,Klitzing:2007,Heathcote:2008,Morizot:2008,Sherlock:2011,Merloti:2013}. Using this approach, atoms have been trapped in different geometries which include the double well \cite{Schumm:2005}, quasi-two dimensional traps \cite{Colombe:2004,White:2006,Merloti:2013} and ring traps \cite{Heathcote:2008,Sherlock:2011}.\\
\hspace*{15pt}
The trapping geometries with controlled rotation of atom cloud on a circular path (\textit{i.e.} ring shape) are of considerable interest for several purposes which include the study of super-fluidity in low-dimensional systems \cite{Sherlock:2011}, measurement of quantum mechanical phases \cite{Sherlock:2011,Javanainen:1996}, and the realization of a cold atom gyroscope \cite{Muller:2009} for high precision measurement of rotation. In this work, it is theoretically shown that, using the static magnetic field of a quadrupole trap and RF fields of different polarizations and phases, some novel two-dimensional RF-dressed potentials for an atom can be generated with the support of a suitable dipole trap. The generated anisotropic two-dimensional time-dependent potentials may be useful for the controlled rotation and oscillation of the cold atom cloud on the circular path. \\
\hspace*{15pt}
This article is organized as follows. The section \ref{theory} presents the general theoretical background to calculate the adiabatic dressed state potentials for an atom interacting with a RF field in the presence of a static magnetic field. The eigenvalues of the atom-field interaction Hamiltonian, which give the RF-dressed potentials, are calculated for the RF field of any arbitrary polarization in the presence of a static magnetic field of a quadrupole trap. The section \ref{Discussion} describes the well known RF-dressed potentials for different cases of the RF fields whereas section \ref{Results} shows our proposed trapping potentials that can also be achieved through RF dressing. The RF-dressed potentials obtained for different cases are plotted to show the trapping and manipulation geometries achievable for the cold atoms. Finally, the conclusion of our work is presented in section \ref{conclusion}.

\section{Theory}\label{theory}
The origin of the adiabatic dressed state potential due to interaction of an atom with the RF field in the presence of a static magnetic field has been described earlier in several research works \cite{Morizot:2006,Heathcote:2008,Morizot:2008,Lesanovsky:2006:73,Gildemeister:thesis}. It can be understood well within the frame work of semi-classical theory as described in the following. The dressed state potentials presented here have been worked out with example of $ ^{87}Rb $ atoms. \\
\hspace*{15pt}
We assume the static magnetic field due to a quadrupole magnetic trap is given as \cite{Yum:2012},
\begin{equation}\label{eq:quadfield}
\textbf{B}^\textbf{S}\textbf{(r)}=B_q\left(\ x\hat{e}_x+y\hat{e}_y-2z\hat{e}_z \right) = B_q\left(\begin{array}{c}
x\\
y\\
-2z\\
\end{array}\right),
\end{equation}
and time-dependent Rf field  $\textbf{B}^{\textbf{RF}}\textbf{(t)}$ is given as,
\begin{equation}\label{eq:rffieldshort}
\textbf{B}^\textbf{{RF}}\textbf{(t)}=B_x^{RF}(t)\hat{e}_x+B_y^{RF}(t)\hat{e}_y+B_z^{RF}(t)\hat{e}_z,
\end{equation}
where $\hat{e}_x$, $\hat{e}_y$, and $\hat{e}_z$ are the basis unit vectors in the laboratory coordinate system. The parameter $2B_q$ is the quadrupole field gradient along the z-axis. \\
\hspace*{5pt}
Neglecting the effect of gravity and the kinetic energy of the atom, the time dependent Hamiltonian for the atom having hyperfine angular momentum \textbf{F} and interacting with a magnetic field $\textbf{B(r,t)}$ is given by \cite{Lesanovsky:2006:73}
\begin{equation}\label{eq:hamiltonian}
H(t)=-\mu.\textbf{B}=\frac{g_F\mu_B}{\hbar}{\textbf{F}.\textbf{B(r,t)}},
\end{equation}
where magnetic field \textbf{B(r,t)} in this equation is sum of static and time-dependent fields given by Eqs. (\ref{eq:quadfield}) and (\ref{eq:rffieldshort}), and it can be expressed as,
\begin{equation}\label{magfield}
\textbf{B(r,t)}=\textbf{B}^\textbf{S}(\textbf{r})+\textbf{B}^\textbf{{RF}}\textbf{(t)}=|\textbf{B}^\textbf{S}(\textbf{r})|\hat{e}_S+\textbf{B}^\textbf{{RF}}\textbf{(t)},
\end{equation}
with $\hat{e}_S$ being an unit vector along the direction of the static field.\\
\hspace*{15pt}
In order to evaluate adiabatic potentials for an atom in the magnetic field given by Eq. (\ref{magfield}), we follow the theoretical approach as described earlier \cite{Gildemeister:thesis}. In this approach, first, the Hamiltonian of Eq. (\ref{eq:hamiltonian}) is evaluated in a new coordinate system whose one basis vector is parallel to the local magnetic field direction. In this new local coordinate system, represented by unit basis vectors $(\hat{e}_1,\hat{e}_2,\hat{e}_3)$, the unit vector $\hat{e}_3$ is along the direction of local static magnetic field ($\hat{e}_S$), and the other two unit vectors (\textit{i.e.} $\hat{e}_1$ and $\hat{e}_2$) are perpendicular to $\hat{e}_s$ as well as to each other. Due to variation of $\textbf{B}^\textbf{S}\textbf{(r)}$ in space, these unit vectors $(\hat{e}_1,\hat{e}_2,\hat{e}_3)$ turn out to be position dependent when expressed in laboratory coordinate system. In this new local coordinate system, the appearance of static magnetic field of the quadrupole trap is $(0,0,|\textbf{B}^\textbf{S}(\textbf{r})|)$ and that of RF field is $(B_{\bot 1}^{RF}(t), B_{\bot 2}^{RF}(t), B_{||}^{RF}(t))$. The Hamiltonian in Eq. (\ref {eq:hamiltonian}) in this local coordinate system is given as,
\begin{equation}
H(t)=\frac{g_F\mu_B}{\hbar}\left[F_3|\textbf{B}^\textbf{S}|+F_3B^{RF}_{||}(t)+F_1B^{RF}_{\bot 1}(t)+F_2B^{RF}_{\bot 2}(t)\right].
\end{equation}
Here $F_1$, $F_2$ and $F_3$ are components of $\textbf{F}$ along $\hat{e}_1, \hat{e}_2$ and $\hat{e}_3$ axes. Since $\hat{e}_3$ is the unit vector along the vector $\hat{e}_S$ (\textit{i.e.} the direction of the local static field), it represents the local quantization axis.\\
\hspace*{15pt}
In the next step the Hamiltonian is transformed to a frame rotating about $F_3$ by applying the unitary transformation $U=e^{\frac{i}{\hbar}(\omega tF_3)}$, which modifies the state $|\psi\rangle$ to $|\psi\rangle^R$ as
\begin{equation}
|\psi\rangle^R=U|\psi\rangle,
\end{equation}
and the Schrodinger equation as
\begin{equation}
H_R|\psi\rangle^R=i\hbar\frac{\partial |\psi\rangle^R}{\partial t},
\end{equation}
where
\begin{equation}\label{eq:HRt}
H_R(t)=UHU^\dagger+i\hbar\frac{\partial U}{\partial t}U^\dagger.
\end{equation}
We can write the time variation of RF field as $B_{\bot 1}^{RF}(t)=B_{\bot 1}^{RF}\cos(\omega t)$ , $B_{\bot 2}^{RF}(t)=B_{\bot 2}^{RF}\cos(\omega t+\gamma)$ and $B_{||}^{RF}(t)=B_{||}^{RF}\cos(\omega t+\gamma')$, where $\gamma$ and $\gamma'$ are the phases of the RF field along the $\hat{e}_2$ and $\hat{e}_3$ unit vectors. After applying the rotating-wave approximation, we get the time independent Hamiltonian using Eq. (\ref{eq:HRt}) as,
\begin{widetext}
\begin{multline}\label{eq:HRti}
H_R(t)=H_R\\=\frac{g_F\mu_B}{\hbar}\left[F_3\left(|\textbf{B}^\textbf{S}|-\frac{\hbar\omega}{g_F\mu_B}\right)+\left\{\frac{F_+}{4}(B^{RF}_{\bot 1}-iB^{RF}_{\bot 2} e^{i\gamma})+\frac{F_-}{4}(B^{RF}_{\bot 1}+iB^{RF}_{\bot 2} e^{-i\gamma})\right\}\right].
\end{multline}
\end{widetext}
Here $F_+=F_1+iF_2$ and $F_-=F_1-iF_2$ are the raising and lowering operator for the hyperfine angular momentum respectively. For an atom in $|F,m_F\rangle$ manifold with F=2 and $m_F=(2,1,0,-1,-2)$, the interaction Hamiltonian can be written in matrix form as,
\begingroup
\fontsize{8pt}{8pt}
\begin{widetext}
\begin{equation}\label{eq:HR}
\frac{H_R}{g_F\mu_B}=\begin{bmatrix}
2(|B^S|-\frac{\hbar\omega}{g_F\mu_B})         &\frac{1}{2}(B^{RF}_{\bot 1}-iB^{RF}_{\bot 2} e^{i\gamma}) &0 &0 &0\\
\frac{1}{2}(B^{RF}_{\bot 1}+iB^{RF}_{\bot 2} e^{-i\gamma})        &(|B^S|-\frac{\hbar\omega}{g_F\mu_B})    &\frac{\sqrt{6}}{4}(B^{RF}_{\bot 1}-iB^{RF}_{\bot 2} e^{i\gamma}) &0 &0\\
0   &\frac{\sqrt{6}}{4}(B^{RF}_{\bot 1}+iB^{RF}_{\bot 2} e^{-i\gamma}) &0   &\frac{\sqrt{6}}{4}(B^{RF}_{\bot 1}-iB^{RF}_{\bot 2} e^{i\gamma}) &0\\
0   &0    &\frac{\sqrt{6}}{4}(B^{RF}_{\bot 1}+iB^{RF}_{\bot 2} e^{-i\gamma})   &-(|B^S|-\frac{\hbar\omega}{g_F\mu_B})   &\frac{1}{2}(B^{RF}_{\bot 1}-iB^{RF}_{\bot 2} e^{i\gamma})\\
0    &0     &0     &\frac{1}{2}(B^{RF}_{\bot 1}+iB^{RF}_{\bot 2} e^{-i\gamma})   &-2(|B^S|-\frac{\hbar\omega}{g_F\mu_B}) \\
\end{bmatrix}.
\end{equation}
\end{widetext}
\endgroup
The diagonalization of the matrix in Eq. (\ref{eq:HR}) gives the energy eigenvalues of the Hamiltonian $H_R$ in Eq. (\ref{eq:HRti}). These  eigenvalues give the adiabatic dressed state potentials (V) for the atom-radiation system. After diagonalizing the matrix in Eq. (\ref{eq:HR}), the dressed state potential for Zeeman sub-level $|F,m_F\rangle$ can be written as,
\begin{widetext}
\begin{equation}\label{eq:potential}
V=m_Fg_F\mu_B\sqrt{\left(|\textbf{B}^\textbf{S}|-\frac{\hbar\omega}{g_F\mu_B}\right)^2+\frac{1}{4}\left[(B^{RF}_{\bot 1})^2+(B^{RF}_{\bot 2})^2+2B^{RF}_{\bot 1}B^{RF}_{\bot 2}\sin\gamma\right]} .
\end{equation}
\end{widetext}
Now identifying the detuning ($\delta$) and Rabi frequency ($\Omega$) of the RF field by
\begin{equation}\label{eq:delta}
\delta=\omega-\frac{g_F\mu_B}{\hbar}|\textbf{B}^\textbf{S}|,
\end{equation}
and
\begin{equation}\label{eq:omega}
|\Omega|^2=\left(\frac{g_F\mu_B}{2\hbar}\right)^2{\left[(B^{RF}_{\bot 1})^2+(B^{RF}_{\bot 2})^2+2B^{RF}_{\bot 1} B^{RF}_{\bot 2}\sin\gamma\right]},
\end{equation}
we can rewrite Eq. (\ref{eq:potential}) as,
\begin{equation}\label{eq:pot_compact}
V=m_F\hbar\sqrt{\delta^2+|\Omega|^2}.
\end{equation}
It is important to note that Rabi frequency given by Eq. (\ref{eq:omega}) does not have contribution from the parallel component of the RF field $B_{||}^{RF}(t)$. Thus, components of the RF field perpendicular to the local static field (\textit{i.e.} $B_{\bot 1}^{RF}(t)$ and $B_{\bot 2}^{RF}(t)$) are only important for the evaluation of Rabi frequency. For a time varying RF field at frequency $\omega$ with its x-, y- and z-components in the laboratory coordinate system described by
\begin{equation}\label{eq:rffield}
\textbf{B}^\textbf{{RF}}\textbf{(t)}=\left(\begin{array}{c}
B_x^{RF}(t)\\
B_y^{RF}(t)\\
B_z^{RF}(t)\\
\end{array}\right)
=\left(\begin{array}{c}
B_x\cos\omega t\\
B_y\cos(\omega t-\alpha)\\
B_z\cos(\omega t-\beta)\\
\end{array}\right),
\end{equation}
the calculation of the Rabi frequency involves the calculation of $B_{\bot 1}^{RF}(t)$ and $B_{\bot 2}^{RF}(t)$. This is briefly discussed as follows. \\
\hspace*{15pt}
In the laboratory coordinate system, the unit vectors along x-, y-, and z-axes (\textit{i.e.} $\hat{e}_x$, $\hat{e}_y$, and $\hat{e}_z$) can be expressed in the column vector form as,
\begin{eqnarray}
\hat{e}_x=\left(\begin{array}{c}
1\\
0\\
0\\
\end{array}\right),\
\hat{e}_y=\left(\begin{array}{c}
0\\
1\\
0\\
\end{array}\right),\
\hat{e}_z=\left(\begin{array}{c}
0\\
0\\
1\\
\end{array}\right).
\end{eqnarray}
\hspace*{5pt}
For the quadrupole magnetic trap with its static magnetic field given by the Eq. (\ref{eq:quadfield}), the basis vectors of the local coordinate system (\textit{i.e.} $\hat{e}_1$, $\hat{e}_2$ and $\hat{e}_3$) can also be expressed in as column vectors in the laboratory coordinate system as \cite{Heathcote:thesis},
\begin{eqnarray}
\hat{e}_1=\left(\begin{array}{c}
cos\theta cos\phi\\
cos\theta sin\phi\\
sin\theta\\
\end{array}\right),\
\hat{e}_2=\left(\begin{array}{c}
sin\phi\\
-cos\phi\\
0\\
\end{array}\right),\
\hat{e}_3=\left(\begin{array}{c}
sin\theta cos\phi\\
sin\theta sin\phi\\
-cos\theta\\
\end{array}\right),
\end{eqnarray}
where $\cos\phi=\frac{x}{\sqrt{x^2+y^2}}$, $\sin\phi=\frac{y}{\sqrt{x^2+y^2}}$, $\cos\theta=\frac{2z}{\sqrt{x^2+y^2+4z^2}}$, and $\sin\theta=\frac{\sqrt{x^2+y^2}}{\sqrt{x^2+y^2+4z^2}}$.\\
\hspace*{15pt}
The rotation matrices $\Re(\theta,\phi)$ and $\Re^{-1}(\theta,\phi)$ defined by 
\begin{equation}\label{rotmat}
\Re(\theta,\phi)=\left( \begin{array} {ccc}
\cos\theta\cos\phi  & \sin\phi       & \sin\theta\cos\phi \\
\cos\theta\sin\phi  & -\cos\phi      & \sin\theta\sin\phi \\
\sin\theta          & 0             & -\cos\theta        \\
\end{array}\right)
\end{equation}
and
\begin{equation}\label{invrotmat}
\Re^{-1}(\theta,\phi)=\left( \begin{array} {ccc}
\cos\theta\cos\phi  & \cos\theta\sin\phi & \sin\theta \\
sin\phi             & -cos\phi           & 0          \\
\sin\theta\cos\phi  & \sin\theta\sin\phi & -\cos\theta \\
\end{array}\right)
\end{equation}
can be used to relate these basis vectors $(\hat{e}_1,\hat{e}_2,\hat{e}_3)$ to the basis vectors $(\hat{e}_x,\hat{e}_y,\hat{e}_z)$ as following:
$\Re(\theta,\phi) \hat{e}_x = \hat{e}_1$, $\Re(\theta,\phi) \hat{e}_y=\hat{e}_2$ and $\Re(\theta,\phi) \hat{e}_z=\hat{e}_3$;
$\Re^{-1}(\theta,\phi)\hat{e}_1=\hat{e}_x$, $\Re^{-1}(\theta,\phi)\hat{e}_2=\hat{e}_y$ and $\Re^{-1}(\theta,\phi)\hat{e}_3=\hat{e}_z$.
Using the above transformation matrices $\Re(\theta,\phi)$ or $\Re^{-1}(\theta,\phi)$, any arbitrary vector (such as applied RF magnetic field) expressed in one basis set (either $(\hat{e}_x,\hat{e}_y,\hat{e}_z)$ or $(\hat{e}_1,\hat{e}_2,\hat{e}_3)$) can be transformed into a vector expressed in the other basis set. Thus the RF magnetic field vector, which is given in laboratory coordinate system by Eq. (\ref{eq:rffield}), is transformed to a vector $\Re^{-1}(\theta,\phi)\textbf{B}^{\textbf{RF}}\textbf{(t)}$ in the local coordinate system. The vector $\Re^{-1}(\theta,\phi)\textbf{B}^{\textbf{RF}}\textbf{(t)}$ is given as,
\begin{widetext}
\begin{multline}\label{eq:rotmatrf}
\Re^{-1}(\theta,\phi)\textbf{B}^\textbf{{RF}}\textbf{(t)}\\=\left(\begin{array}{c}
B_x\cos\theta\cos\phi\cos\omega t+B_y\cos\theta\sin\phi\cos(\omega t-\alpha)+B_z\sin\theta\cos(\omega t-\beta)\\
B_x\sin\phi\cos\omega t-B_y\cos\phi\cos(\omega t-\alpha)\\
B_x\sin\theta\cos\phi\cos\omega t+B_y\sin\theta\sin\phi\cos(\omega t-\alpha)-B_z\cos\theta\cos(\omega t-\beta)\\
\end{array}\right).
\end{multline}
\end{widetext}
Evidently, the first and second elements of the column vector in Eq. (\ref{eq:rotmatrf}) give $B_{\bot 1}^{RF}(t)$ and $B_{\bot 2}^{RF}(t)$, which are required to evaluate Rabi frequency using Eq. (\ref{eq:omega}). The third element of the column vector $\Re^{-1}(\theta,\phi)\textbf{B}^{\textbf{RF}}\textbf{(t)}$  represents RF field along vector $\hat{e_3}$. Hence using Eq. (\ref{eq:omega}) and Eq.  (\ref{eq:rotmatrf}), one gets the expression for the Rabi frequency as 
\begin{widetext}
\begin{multline}\label{eq:omegafinal}
|\Omega|^2=\left(\frac{g_F\mu_B}{2\hbar}\right)^2 \Bigg[\frac{4z^2}{x^2+y^2+4z^2}\left( \frac{B_x^2x^2+B_y^2y^2}{x^2+y^2}\right) +\left(\frac{B_x^2y^2+B_y^2x^2}{x^2+y^2}\right)\\+B_z^2\left( \frac{x^2+y^2}{x^2+y^2+4z^2}\right)-\frac{2B_xB_yxy\cos\alpha}{x^2+y^2+4z^2}+\frac{4B_xB_yz\sin\alpha}{\sqrt{x^2+y^2+4z^2}}+\frac{4B_yB_zyz\cos(\alpha-\beta)}{x^2+y^2+4z^2}\\+\frac{2B_yB_zx\sin(\alpha-\beta)}{\sqrt{x^2+y^2+4z^2}}+\frac{4B_zB_xzx\cos\beta}{x^2+y^2+4z^2}+\frac{2B_zB_xy\sin\beta}{\sqrt{x^2+y^2+4z^2}}\Bigg].
\end{multline}
\end{widetext}
Using this expression for Rabi frequency $|\Omega|$ (\textit{i.e.} Eq. (\ref{eq:omegafinal})), the dressed state potential (Eq. (\ref{eq:pot_compact})) for an atom, interacting with the RF field and static quadrupole trap field, can be obtained.

\section{Discussion}\label{Discussion}
\hspace*{5pt}
The Eqs. (\ref{eq:pot_compact}) and (\ref{eq:omegafinal}) provide a general expression for the RF-dressed potential (V) for an atom in a Zeeman hyperfine sub-level $|F,m_F\rangle$ interacting with a static magnetic field $\textbf{B}^\textbf{S}\textbf{(r)}$ (given by Eq. (\ref{eq:quadfield})) and a RF field $\textbf{B}^{\textbf{RF}}\textbf{(t)}$ given by Eq. (\ref{eq:rffield}). In this section, the formalism developed in the previous section has been utilized to obtain the RF-dressed potentials for selected configurations of RF field and static magnetic field of a quadrupole trap. Based on this, different geometries for trapping and manipulation of cold atoms have been discussed, with a particular example of $^{87}Rb$ atom in $|F=2,m_F=2\rangle$ state. \\
\hspace*{5pt}
The quadrupole trap suffers from Majorana losses which occur at the trap centre (\textit{i.e.} location of zero magnetic field). To avoid these losses, one can try to trap cold atoms at a plane $z=-z_0$ (away from the centre of quadrupole trap) using some external potential, such as potential of a sheet type red-detuned laser beam. This can confine atoms in xy-plane at $z=-z_0$ \cite{Morizot:2006}. The sheet potential with a tight confinement along vertical direction ( \textit{i.e.} z-direction) results in two-dimensional (2D) motion of atoms in xy-plane. The vertical confinement can also be made strong enough to hold atoms against gravity. At this shifted position, the magnitude of the field is given as
\begin{equation}\label{eq:modfield}
\left|\textbf{B}^\textbf{S}(\textbf{r})\right|_{z=-z_0}=2B_qz_0\sqrt{1+\frac{x^2+y^2}{4z_0^2}}=B_0^S\sqrt{1+\frac{x^2+y^2}{4z_0^2}},
\end{equation}
where $B_0^S=2B_qz_0$.\\
\hspace*{5pt}
A number of 2D trapping geometries can be realized by considering the various forms of RF field in the presence of static magnetic field of the quadrupole trap given by Eq. (\ref{eq:modfield}). For example, a double-well potential can be generated by taking a linearly polarized RF field along x-axis, \textit{i.e.} $\textbf{B}^{\textbf{RF}}\textbf{(t)}=(B_x\cos\omega t,0,0)$, in the presence of above static field. The double-well type RF-dressed potential for an atom in the state $|F,m_F\rangle$ can be evaluated from Eqs. (\ref{eq:pot_compact}) and (\ref{eq:omegafinal}) as 
\begin{equation}\label{eq:dwpot}
V_{DW}=m_F\hbar\sqrt{\delta_0^2+\left(\frac{g_F\mu_B}{2\hbar}\right)^2\left[\frac{B_x^2}{x^2+y^2}\left(\frac{4x^2z_0^2}{x^2+y^2+4z_0^2}+y^2\right)\right]},
\end{equation}
where
\begin{equation}\label{eq:delta0}
\delta_0=\omega-\frac{g_F\mu_BB_0^S}{\hbar}\sqrt{1+\frac{x^2+y^2}{4z_0^2}}.
\end{equation}

\begin{figure}[b]
\includegraphics[width=8.6cm]{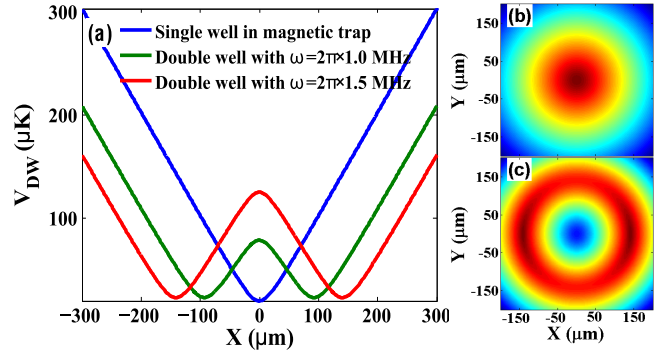}
\caption{\label{fig:doublewell}(Color online) RF-dressed adiabatic potential (Eq. (\ref{eq:dwpot})) showing double-well trapping geometry for the $^{87}Rb$ atom in the state $|F=2,m_F=2\rangle$, for different parameters as $B_q$=150 G/cm, $B_x$=0.7 G, $\omega=2\pi\times 1.5$ MHz and $z_0=$10 $\mu m$. Plot (a) shows the variation in the potential with position along x-axis without RF field (single well) as well as with RF field (double well) for two different values of frequencies ($\omega=2\pi\times$1.0 MHz and $\omega=2\pi\times$1.5 MHz) of RF field. The graphs (b) and (c) show the iso-potential contours in the xy-plane for the single well and double well potentials ($\omega=2\pi\times$1.5 MHz) respectively. The color contrast in contours reveal the variation in potential from minimum (red) to maximum (blue) values through intermediate values shown by orange, yellow and green. }
\end{figure}

Fig. \ref{fig:doublewell} shows the plot of double well potential given by Eq. (\ref{eq:dwpot}). By varying the quadrupole field gradient $B_q$ as well as the frequency $\omega$ of the RF field, the separation between the wells (d) can be varied. The relation that governs this dependence can be extracted from the criterion of the potential minimum of Eq. (\ref{eq:dwpot}). It is given as
\begin{equation}
d=4z_0\sqrt{\left(\frac{\hbar\omega}{2g_F\mu_B|z_0|B_q}\right)^2-1}.
\end{equation}
\hspace*{5pt}
Figure \ref{fig:doublewell} shows the double well potential along the x-axis, whereas it is obvious that with a y-polarized RF field, the double well potential along the y-axis can be generated. The depth of the potential can be changed by changing the amplitude of the RF field (\textit{i.e.} $B_x$ or $B_y$). The generation of such type of a double well potentials has been experimentally realized using a RF field and a Ioffe-Pritchard trap \cite{Lesanovsky:2006:73}. The double well potential as a beam splitter for a Bose-Einstein condensate has also been demonstrated in a miniaturized trap \cite{Schumm:2005}. This double well as a beam splitter has advantage over the static magnetic field beam splitter due to its robustness and coherence preserving properties. Further, it can be shown that incorporating a RF field along z-axis (i.e. $B_z$), the symmetric double well potential can be converted to asymmetric double well potential with unequal depth of two wells. Such type of potential can be useful for unequal splitting of atom cloud as well as for tunnelling studies \cite{Albiez:2005}.\\
\begin{figure}[b]
\includegraphics[width=8.6cm]{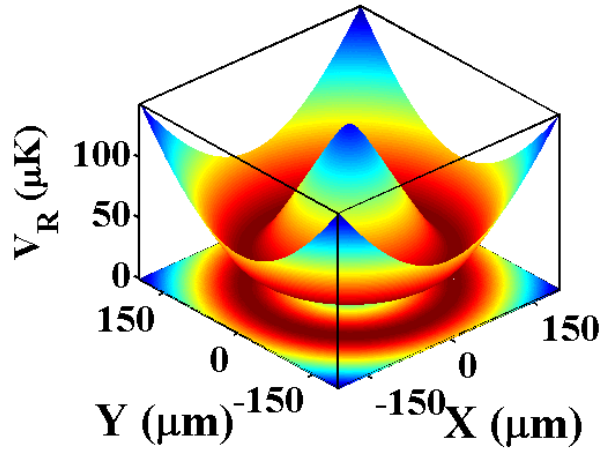}
\caption{\label{fig:ring}(Color online) The ring shaped trapping geometry with the RF-dressed potential of Eq. (\ref{eq:ringpot}) for the $^{87}Rb$ atom in the state $|F=2,m_F=2\rangle$. The trapping parameters are $B_q$=150 G/cm, $B_x$=$B_y$=0.7 G, $\omega=2\pi\times 1.5$ MHz and $z_0=$10 $\mu m$. The elevated plot shows the actual variation of the potential with x and y coordinates, whereas the carpet shows the iso-potential contours in the xy-plane at $z_0=$10 $\mu m$. The color contrast in the contours reveal the variation in potential from minimum (red) to maximum (blue) values through the intermediate values shown by orange, yellow and green colors. The ring radius is nearly 140 $\mu m$.}
\end{figure}
\hspace*{5pt}
Similarly, a well known trapping geometry with RF-dressed potentials is a ring shaped trapping potential \cite{Klitzing:2007,Heathcote:2008,Sherlock:2011}. The ring shaped trapping potential can be generated by using a circularly polarized RF field of the form $\textbf{B}^{\textbf{RF}}(\textbf{t})=(B_x\cos\omega t, B_y\cos(\omega t+\frac{\pi}{2}), 0)$, with $B_x=B_y$, and static magnetic field given by Eq. (\ref{eq:modfield}). For these fields, the Eqs. (\ref{eq:pot_compact}) and (\ref{eq:omegafinal}) evidently result the RF-dressed potential (for an atom in the state $|F,m_F\rangle$) with spatial variation given as, 
\begin{equation}\label{eq:ringpot}
V_R=m_F\hbar\sqrt{\delta_0^2+B_x^2\left(\frac{g_F\mu_B}{2\hbar}\right)^2\left[\frac{4z_0^2}{x^2+y^2+4z_0^2}+\frac{4z_0}{\sqrt{x^2+y^2+4z_0^2}}+1\right]},
\end{equation}
where $\delta_0$ is given by Eq. (\ref{eq:delta0}). The potential given by Eq. (\ref{eq:ringpot}) is independent of polar angle $\phi$ and its plot in the xy-plane is shown in Fig. \ref{fig:ring}. From the graph, it is evident that minimum of the potential lies on a circular ring in the xy-plane. The radius (R) of this ring can be obtained \cite{Heathcote:2008} as 
\begin{equation}\label{eq:radius}
R=2z_0\sqrt{\left(\frac{\hbar\omega}{g_F\mu_BB_0^S}\right)^2-1}.
\end{equation}
\begin{figure}
\includegraphics[width=8.6cm]{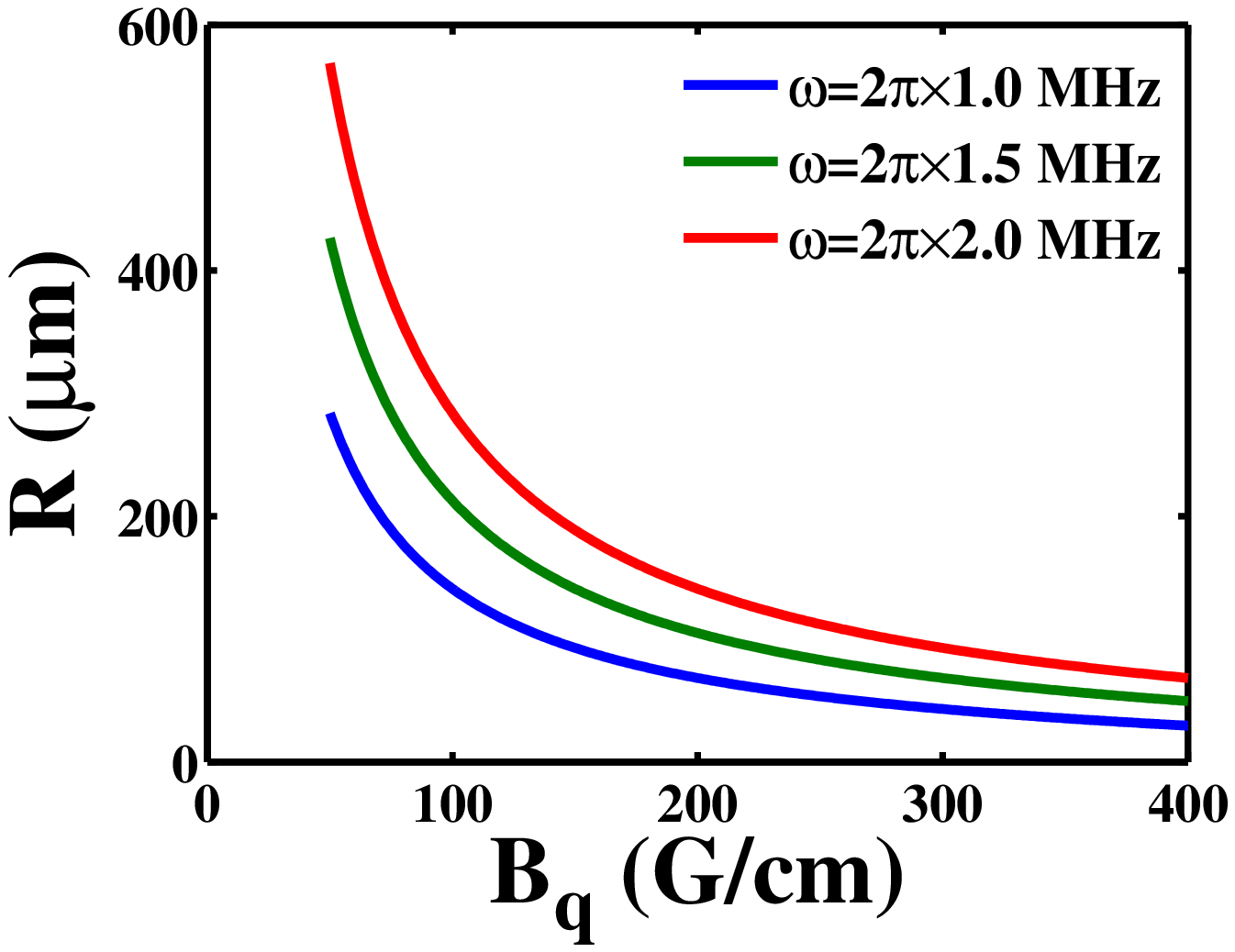}
\caption{\label{fig:ringradius}(Color online) Variation in the ring radius (R) in the ring shaped trapping potential with the quadrupole field gradient $B_q$ for different frequencies of RF field, for $^{87}Rb$ atom in $|F=2,m_F=2\rangle$ state. The other parameters are $B_x=B_y$=0.7 G and $z_0=$10 $\mu m$.}
\end{figure}
\hspace*{5pt}
From the expression of the ring radius (\textit{i.e.} Eq. (\ref{eq:radius})), it is obvious that the radius of the ring trap can be varied by varying $\omega$ as well as $B_q$. The Fig. \ref{fig:ringradius} shows the variation of this radius with $B_q$ for different values of $\omega$. It is preferable to change the radius by varying the quadrupole field gradient only, rather than by sweeping the frequency $\omega$ \cite{Sherlock:2011}. During the frequency sweep, the non-adiabatic Landau-Zener transitions may become significant and increase the loss of trapped atoms due to transition of atoms to untrappable states. Alternatively, the radius of ring can be varied by varying the position of trapping plane ($Z_0$) at a fixed $B_q$ \cite{Heathcote:2008}.\\ 
\hspace*{5pt}
Further, instead of a circularly polarized RF field, if an elliptically polarized RF field is chosen, the generated potential is split arcs. With a RF field of the form $\textbf{B}^{\textbf{RF}}\textbf{(t)}=(B_x\cos\omega t, B_y\cos(\omega t+\frac{\pi}{2}), 0)$, with $B_x \ne B_y$, and static field as given by Eq. (\ref{eq:modfield}), the RF-dressed potential for split arcs trapping geometry is given as,
\begin{figure}[t]
\includegraphics[width=8.6cm]{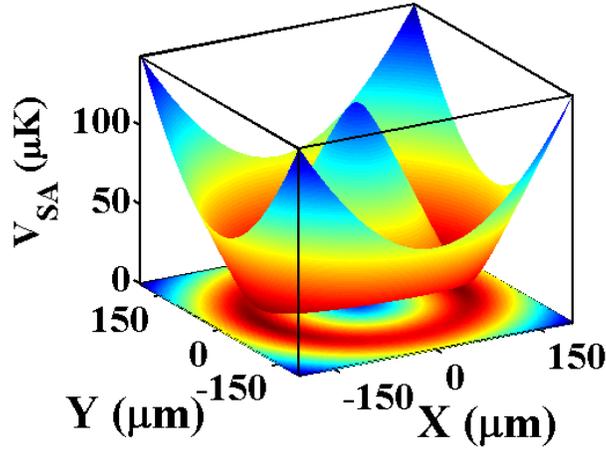}
\caption{\label{fig:arc}(Color online) The plots of RF-dressed potential of Eq. (\ref{eq:splitarcpot}) showing the split arcs trapping potential for the $^{87}Rb$ atom in the state $|F=2,m_F=2\rangle$. The other parameters are $B_q$=150 G/cm, $B_x$=0.7 G, $B_y$=0.14 G, $\omega=2\pi\times 1.5$ MHz and $z_0=$10 $\mu m$. The elevated plot shows the actual variation of the potential whereas the carpet shows the iso-potential contours. The color contrast in contours reveal the variation in potential from minimum (red) to maximum (blue) values through intermediate values shown by orange, yellow and green.}
\end{figure}
\begin{equation}\label{eq:splitarcpot}
V_{SA}=m_F\hbar\sqrt{\delta_0^2+|\Omega|_{SA}^2},
\end{equation}
where
\begin{equation}
|\Omega|_{SA}^2=\left(\frac{g_F\mu_B}{2\hbar}\right)^2 \left[\frac{4z_0^2}{x^2+y^2+4z_0^2}\left(\frac{B_x^2x^2+B_y^2y^2}{x^2+y^2}\right)\\ +\left(\frac{B_x^2y^2+B_y^2x^2}{x^2+y^2}\right)+\frac{4B_xB_yz_0}{\sqrt{x^2+y^2+4z_0^2}}\right].
\end{equation}
The plot of Eq. (\ref{eq:splitarcpot}) is shown in the Fig. \ref{fig:arc} for unequal values of $B_x$ and $B_y$. It can be pointed out that the orientation of the arcs is dependent on the values of $B_x$, $B_y$ and phase ($\alpha$). Heathcote \textit{et. al.} \cite{Heathcote:2008} experimentally demonstrated the variation in orientation of split arcs by varying the relative phase between the components of the RF field.\\
\hspace*{5pt}
The ring trap for cold atomic gases are promising for studying low dimensional systems where one or more dimensions are quantum mechanically restricted for motion. The ring trap geometry can also be useful for cold atom based inertial sensors such as atom gyroscope \cite{Sauer:2001}. 

\section{Results}\label{Results}
In this section, we discuss our proposed two-dimensional trapping potentials that can be generated with appropriate choice of RF field and its phase. These potentials described in the following may be useful for various purposes.

\subsection{Asymmetric split arcs potential}
It is expected that split arcs with unequal depth of trapping potential may be useful to study quantum mechanical tunneling with cold atoms trapped in the arcs. We note that such arcs with unequal potential depth can be created by incorporating the z-component in the RF field. Thus, if the RF field is taken as $\textbf{B}^{\textbf{RF}}(\textbf{t})=(B_x\cos\omega t, B_y\cos(\omega t+\frac{\pi}{2}), B_z\cos(\omega t-\beta))$, with $B_x \ne B_y$, and static field as given by Eq. (\ref{eq:modfield}), the RF-dressed potential can be written as 
\begin{equation}\label{eq:pot_asy_split_arcs}
V_{ASA}=m_F\hbar\sqrt{\delta_0^2+|\Omega|_{ASA}^2},
\end{equation}
where
\begin{widetext}
\begin{multline}\label{eq:omegat_asy_split_arcs}
|\Omega|_{ASA}^2=\left(\frac{g_F\mu_B}{2\hbar}\right)^2 \Bigg[\frac{4z_0^2}{x^2+y^2+4z_0^2}\left( \frac{B_x^2x^2+B_y^2y^2}{x^2+y^2}\right) +\left(\frac{B_x^2y^2+B_y^2x^2}{x^2+y^2}\right)\\+B_z^2\left(\frac{x^2+y^2}{x^2+y^2+4z_0^2}\right)+\frac{4B_xB_yz_0}{\sqrt{x^2+y^2+4z_0^2}}+\frac{4B_yB_zyz_0\sin\beta}{x^2+y^2+4z_0^2}\\-\frac{2B_yB_zx\cos\beta}{\sqrt{x^2+y^2+4z_0^2}}-\frac{4B_zB_xz_0x\cos\beta}{x^2+y^2+4z_0^2}+\frac{2B_zB_xy\sin\beta}{\sqrt{x^2+y^2+4z_0^2}}\Bigg].
\end{multline}
\end{widetext}
\begin{figure}[h]
\includegraphics[width=8.6cm]{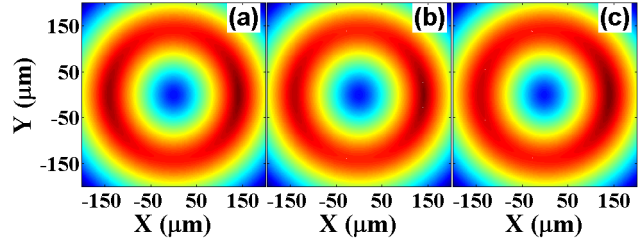}
\caption{\label{fig:assymetric_split_field}(Color online) The iso-potential contours of Eq. (\ref{eq:pot_asy_split_arcs}) showing the asymmetric split arcs for different values of $B_z$: (a) $B_z$=0.07 G, (b) $B_z$=0.14 G, and (c) $B_z$=0.35 G. The other parameters are $B_q$=150 G/cm, $B_x$=0.7 G, $B_y$=0.07 G, $\beta=$0, $\omega=2\pi\times 1.5$ MHz, $z_0=$10 $\mu m$, and $^{87}Rb$ atom in $|F=2,m_F=2\rangle$ state. The color contrast in contours reveal the variation in potential from minimum (red) to maximum (blue) values through intermediate values shown by orange, yellow and green.}
\end{figure}
\begin{figure}[h]
\includegraphics[width=8.6cm]{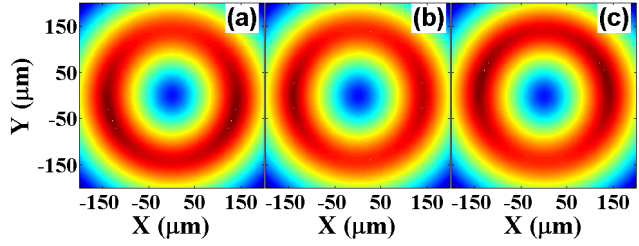}
\caption{\label{fig:assymetric_split_angle}(Color online) The iso-potential contours of Eq. (\ref{eq:pot_asy_split_arcs}) showing the asymmetric split arcs for different values of the phase angles $\beta$: (a) $\beta$=$\frac{\pi}{2}$, (b) $\beta$=$\pi$, and (c) $\beta$=$\frac{3\pi}{2}$. The other parameters are $B_q$=150 G/cm, $B_x$=0.7 G, $B_y$=0.07 G, $B_z=$0.14 G, $\omega=2\pi\times 1.5$ MHz, $z_0=$10 $\mu m$, and $^{87}Rb$ atom in $|F=2,m_F=2\rangle$ state. The color contrast in contours reveal the variation in potential from minimum (red) to maximum (blue) values through intermediate values shown by orange, yellow and green.}
\end{figure}
The contours of the RF-dressed potentials given by Eq. (\ref{eq:pot_asy_split_arcs}) for different values of $B_z$ and $\beta$ are shown in Fig. \ref{fig:assymetric_split_field} and Fig. \ref{fig:assymetric_split_angle} respectively. These graphs clearly reveal that split arcs become asymmetric by incorporation of z-component in the RF field. The ultracold atoms trapped in these split arcs potentials may be a good tool to study tunnelling and Josephson oscillations \cite{Bloch:1973,Albiez:2005} with tunnelling along circular path. As described in Ref. \cite{Albiez:2005}, the asymmetry of the two trapping potentials can be used to control Josephson Oscillations of population between trapping potentials. By increasing the asymmetry, the Josephson oscillations can be suppressed and self trapping can be reached. In our proposed asymmetric split arc potential, the asymmetry can be controlled by adjusting the amplitude and phase of third RF component of the field. The separation between the minima of arcs depends upon the arc length which can be controlled by controlling the radius of the ring as well as the phase angle $\beta$. The $\beta$ dependent separation between potential minima along the arcs is shown in Fig. \ref{fig:assymetric_split_angle}.

\subsection{Asymmetric ring potential}
\hspace*{5pt}
An asymmetric ring potential can be generated by the use of a RF field of the form $\textbf{B}^{\textbf{RF}}(\textbf{t})=(B_x\cos\omega t, B_y\cos(\omega t+\frac{\pi}{2}), B_z\cos(\omega t-\beta))$, with $B_x=B_y$, in the presence of the static magnetic field given by Eq. (\ref{eq:modfield}). The incorporation of z-component in the RF field results in anisotropy in the isotropic ring potential. For these RF and static magnetic fields, the dressed state potential for an atom (\textit{e.g.} $^{87}Rb$) in the state $|F,m_F\rangle$ can be calculated to be 
\begin{equation}\label{eq:pot_tilt_ring}
V_{AR}=m_F\hbar\sqrt{\delta_0^2+|\Omega|_{AR}^2},
\end{equation}
where
\begin{widetext}
\begin{multline}\label{eq:omegatilt}
|\Omega|_{AR}^2=\left(\frac{g_F\mu_B}{2\hbar}\right)^2 \Bigg[B_x^2\left(\frac{4z_0^2}{x^2+y^2+4z_0^2}+\frac{4z_0}{\sqrt{x^2+y^2+4z_0^2}}+1\right)+B_z^2\left( \frac{x^2+y^2}{x^2+y^2+4z_0^2}\right)\\-\frac{2B_xB_z\left(x\cos\beta-y\sin\beta\right)}{\sqrt{x^2+y^2+4z_0^2}}\left(\frac{2z_0}{\sqrt{x^2+y^2+4z_0^2}}+1\right)\Bigg].
\end{multline}
\end{widetext}
\begin{figure*}
\includegraphics[width=15cm]{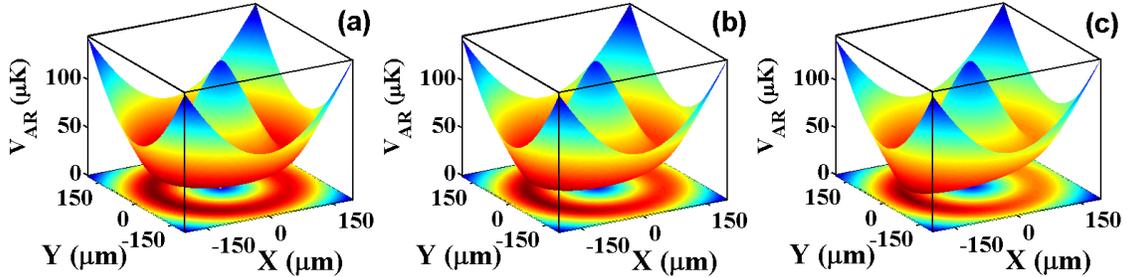}
\caption{\label{fig:tilt}(Color online) The plots of the RF-dressed potential of Eq. (\ref{eq:pot_tilt_ring}) showing the asymmetric ring potential for the $^{87}Rb$ atom in the state $|F=2,m_F=2\rangle$. The graphs shown in the figure are for different values of $B_z$: (a) $B_z$=0.21 G, (b) $B_z$=0.35 G, and (c) $B_z$=0.56 G. The other common parameters used are $B_q$=150 G/cm, $B_x=B_y$=0.7 G, $z_0$=10 $\mu m$, $\beta=0$, and $\omega=2\pi\times 1.5$ MHz. As the value of $B_z$ is increased, from (a) to (c), the tilt in the potential shape is increased due to increase in well depth.}
\end{figure*}
The plots of the potential given by Eqs.(\ref{eq:pot_tilt_ring}) and (\ref{eq:omegatilt}) are shown in Fig. \ref{fig:tilt}, for different values of RF field amplitude $B_z$ at a given value of $\beta$. Due to presence of z-component in the RF field (\textit{i.e.} $B_z$), a potential well is formed at a particular position on the ring. In an earlier work \cite{Sherlock:2011}, the asymmetric ring potential, due to presence of this z-component in RF field, has been experimentally observed in a RF-dressed quadrupole trap. \\
\hspace*{5pt}
The depth of the potential well on the ring is dependent on the value of $B_z$, whereas the position of the well (\textit{i.e.} potential minimum) on the ring is dependent on the phase angle $\beta$. The Fig. \ref{fig:tilt} shows that, as $B_z$ is increased, the ring shaped potential becomes more asymmetric (\textit{i.e.} tilted) with increased well depth. The atoms from the whole ring can be forced to drain to this well position by suitably adjusting the value of $B_z$. This makes the transfer of cold atoms very convenient to a particular site on the ring. By changing the value of $\beta$, the position of potential minimum (\textit{i.e.} the well) can be changed to our choice, which can act as a `conveyor belt' for the cold atom cloud. The transport of the atoms along the ring may also be useful in areas such as atom-interferometry for rotation sensing \cite{Muller:2009}.

\subsection{Rotating potentials}\label{rottrans}
\hspace*{5pt}
As discussed earlier in section \ref{Discussion}, the ring traps provide the confinement to atom cloud from two dimensions which makes the motion of atoms in quasi-one-dimension along the circular path. Such systems may be useful to study non-linear atom interferometry, where confinements of Bose condensates to low dimensions may result in increased non-linear interactions and hence improved measuring sensitivity \cite{Alexander:2010}.\\
\hspace*{5pt}
As discussed before in this article, the inclusion of a z-component in the RF field used for generation of the ring trap geometry results in a potential well on the ring. The atoms from the whole ring can be accumulated in this well. Since the position of this potential well on the ring is dependent on the phase angle $\beta$, by varying the value of $\beta$ with time, the position of the well (and atom cloud trapped in it) can be moved along the circumference of the ring. This rotational motion of the atom cloud seems controllable by controlling the phase modulation in the z-component of the RF field. In the following, we consider two types of phase modulations in the z-component of the RF field. One is the linear phase modulation and the other is the periodic phase modulation. These phase modulations can result in two different types of rotational motion of potential well and atom-cloud trapped in it, which are discussed in the following description.\\
\hspace*{5pt}
The linear phase modulation in the z-component of RF field results in rotation of the potential well on the ring circumference at a rate which is determined by the modulation frequency. For the $ ^{87}Rb$ atoms in $|F,m_F\rangle$ state, first, we consider here the linear phase modulated RF field of the form 
\begin{equation}\label{eq:rflinearphasemod}
\textbf{B}^{\textbf{RF}}\textbf{(t)}=\left(\begin{array}{c}
B_x\cos\omega t\\
B_y\cos(\omega t+\pi/2)\\
B_z\cos(\omega t-\omega_l t)\\
\end{array}\right),
\end{equation}
where the value of modulation frequency $\omega_l$ is very small compared to $\omega$, so that the modified potential due to phase modulation can be adiabatically followed by the atoms. Using the RF field as in Eq. (\ref{eq:rflinearphasemod}) with $B_x=B_y$, and the static field as in Eq. (\ref{eq:modfield}), the Eqs. (\ref{eq:potential}), (\ref{eq:delta}) and (\ref{eq:omegafinal}) result in rotating potential ($V_{RP}$) of the form 
\begin{equation}\label{eq:linearphasepot}
V_{RP}=m_F\hbar\sqrt{\delta_0^2+|\Omega|_{RP}^2},
\end{equation}
where
\begin{widetext}
\begin{multline}\label{eq:omegalinear}
|\Omega|_{RP}^2=\left(\frac{g_F\mu_B}{2\hbar}\right)^2 \Bigg[B_x^2\left(\frac{4z_0^2}{x^2+y^2+4z_0^2}+\frac{4z_0}{\sqrt{x^2+y^2+4z_0^2}}+1\right)+B_z^2\left( \frac{x^2+y^2}{x^2+y^2+4z_0^2}\right)\\-\frac{2B_xB_z\left[x\cos(\omega_lt)-y\sin(\omega_lt)\right]}{\sqrt{x^2+y^2+4z_0^2}}\left(\frac{2z_0}{\sqrt{x^2+y^2+4z_0^2}}+1\right)\Bigg].
\end{multline}
\end{widetext}
\begin{figure}[h]
\includegraphics[width=8.6cm]{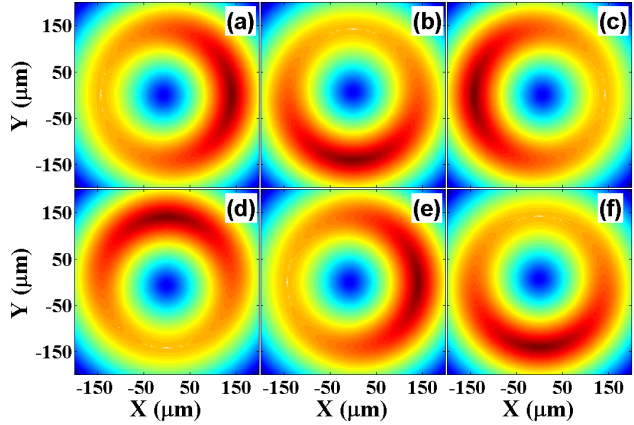}
\caption{\label{fig:rotation}(Color online) The contours of RF-dressed potential of Eq. (\ref{eq:linearphasepot}) for the parameters $B_q$=150 G/cm, $B_x=B_y=B_z=0.7$ G, $\omega=2\pi\times 1.5$ MHz, and $z_0=$10 $\mu m$. The contours are plotted for different values of time t (in seconds): (a) t=$0$ , (b) t=$\frac{\pi}{2\omega_l}$, (c) t=$\frac{\pi}{\omega_l}$, (d) t=$\frac{3\pi}{2\omega_l}$, (e) t=$\frac{2\pi}{\omega_l}$, (f) t=$\frac{5\pi}{2\omega_l}$; with $\omega_l$=1 Hz. The contours show the rotation of potential minimum with time.}
\end{figure}
\hspace*{5pt}
In Eq. (\ref{eq:omegalinear}), the last term in the Rabi frequency expression contains the time dependence, which result the potential of Eq. (\ref{eq:linearphasepot}) to be time dependent such that position of the potential minimum changes with time. The contours in Fig. \ref{fig:rotation} show that  position of the minimum of potential $V_{RP}$ moves along the circular path on the ring with time. Thus, the linear phase modulation discussed above converts the ring potential into a rotating well potential. The rate of rotation is governed by the modulation parameter $\omega_l$. As the rotation takes place, experimentally, it is likely that temperature of the cloud increases due to non-adiabatic following of the potential by atoms in the cloud. This may lower the life-time of the atoms in the trap. However, this rotation provides excellent opportunities to study motion dependent physics phenomenon like super-fluidity in the condensate cloud.\\
\hspace*{15pt}
Next, we consider the periodic phase modulation, in which the phase of the z-component of the RF field is considered to vary periodically. Taking the modulation frequency $\omega_m$, the RF field can be expressed as
\begin{equation}\label{eq:rfperiodicphasemod}
\textbf{B}^{\textbf{RF}}\textbf{(t)}=\left(\begin{array}{c}
B_x\cos\omega t\\
B_y\cos(\omega t+\frac{\pi}{2})\\
B_z\cos(\omega t-\pi\cos(\omega_m t))\\
\end{array}\right).
\end{equation}
Using the RF field as in Eq. (\ref{eq:rfperiodicphasemod}) with $B_x=B_y$, and static field as in Eq. (\ref{eq:modfield}), the Eqs. (\ref{eq:potential}), (\ref{eq:delta}) and (\ref{eq:omegafinal}) result in oscillating potential ($V_{OP}$) (for $^{87}Rb$ atom in $|F,m_F\rangle$ state) of form 
\begin{equation}\label{eq:periodicphasepot}
V_{OP}=m_F\hbar\sqrt{\delta_0^2+|\Omega|_{OP}^2},
\end{equation}
where
\begin{figure}[b]
\includegraphics[width=8.0cm]{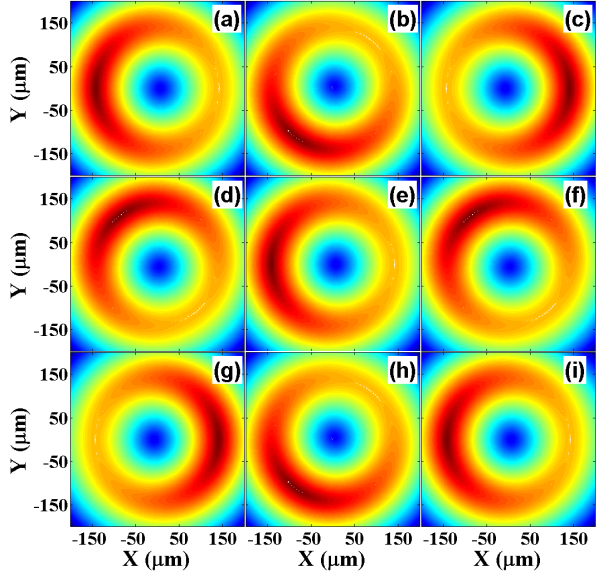}
\caption{\label{fig:minimaintime}(Color online) The contours of RF-dressed potential of Eq. (\ref{eq:periodicphasepot}) for the parameters $B_q$=150 G/cm, $B_x=B_y=B_z=0.7$ G, $\omega=2\pi\times 1.5$ MHz, and $z_0=$10 $\mu m$. The contours are plotted for different values of time t (in seconds): (a) t=$0$, (b) t=$\frac{\pi}{4\omega_m}$, (c) t=$\frac{\pi}{2\omega_m}$, (d) t=$\frac{3\pi}{4\omega_m}$, (e) t=$\frac{\pi}{\omega_m}$, (f) t=$\frac{5\pi}{4\omega_m}$ , (g) t=$\frac{3\pi}{2\omega_m}$ , (h) t=$\frac{7\pi}{4\omega_m}$, (i) t=$\frac{2\pi}{\omega_m}$; with $\omega_m$=1 Hz. The contours show the oscillation of potential minimum with time along the circular path.}
\end{figure}
\begin{widetext}
\begin{multline}\label{eq:omegaperiodic}
|\Omega|_{OP}^2=\left(\frac{g_F\mu_B}{2\hbar}\right)^2 \Bigg[B_x^2\left(\frac{4z_0^2}{x^2+y^2+4z_0^2}+\frac{4z_0}{\sqrt{x^2+y^2+4z_0^2}}+1\right)+B_z^2\left( \frac{x^2+y^2}{x^2+y^2+4z_0^2}\right)\\-\frac{2B_xB_z\left[x\cos(\pi\cos\omega_mt)-y\sin(\pi\cos\omega_mt)\right]}{\sqrt{x^2+y^2+4z_0^2}}\left(\frac{2z_0}{\sqrt{x^2+y^2+4z_0^2}}+1\right)\Bigg].
\end{multline}
\end{widetext}
\hspace*{5pt}
It is evident from the plots of potential given by Eq. (\ref{eq:periodicphasepot}) and Eq. (\ref{eq:omegaperiodic}) shown in Fig. \ref{fig:minimaintime} that the position of minimum of the potential moves along the ring circumference and completes its one rotation in time duration $T_r=\pi/\omega_m$. After a complete rotation in time $T_r$, the sense of rotation gets reversed. The next rotation is completed in same duration $T_r$, but, with sense of rotation in opposite direction. This continues, and the sense of rotation of potential minimum keeps on changing after a time interval of $T_r$. Thus, the atom cloud trapped in this potential makes an oscillatory motion along the circumference of the ring, with a time period of oscillation as $2T_r$.\\
\hspace*{5pt}
The rotating potential described above can be utilized to initiate motion of the Bose condensate ( or cold atom cloud) along the circular path to study super-fluidity, which is otherwise achieved using laser beams of appropriate frequency to transfer momentum from photons to atoms \cite{Ramanathan:2011,Karl:1998,Marques:2001}. The use of laser beams to initiate the rotation of cloud also leads to some heating of the cloud due to finite scattering rate. A rotating cloud in the ring can serve as a singly charged vortex state with vortex core coinciding with centre of the ring. Such singly charged vortices have also been proposed to serve as qubits in quantum information processing \cite{Halkyard:thesis,Kapale:2005}.

\section{Conclusion}\label{conclusion}
\hspace*{5pt}
We have discussed various types of RF-dressed adiabatic potentials for ultracold atoms which can be realized by superimposing an appropriately designed RF field on the static magnetic field of a quadrupole trap. Using this scheme in presence of a supporting two-dimensional dipole trap, apart from the familiar double well and ring trap potentials, some interesting anisotropic (static as well as time-dependent) two-dimensional potentials can be generated with RF fields of various polarizations and phase modulations. The generated time-dependent (i.e. moving) 2D potentials can facilitate the rotation as well as oscillation of the trapped atom cloud along the circular path. Various possible applications of these proposed potentials in transporting atom cloud and in studying Josephson oscillations, super-fluidity and vortex states using Bose condensates in low dimensional geometries have been discussed.

\begin{acknowledgments}
We thank H. S. Rawat, V. B. Tiwari and S.P. Ram for the careful reading of the manuscript and fruitful suggestions. Arijit Chakraborty acknowledges the financial support by the Homi Bhabha National Institute, India.
\end{acknowledgments}


\end{document}